\documentclass[]{emulateapj} \usepackage{apjfonts} \newif\ifdraft \draftfalse \newif\ifsub \subfalse \newif\ifpre \pretrue
\usepackage{graphicx}
\usepackage{epsf}

\shorttitle{The Sixth VLBA Calibrator Survey}
\shortauthors{Petrov et al.}
\received{January~24, 2008}
\accepted{April~30, 2008}
\journalinfo{Astronomical~journal}
\ifpre
  \submitted{}
\else
  \paperid{271422}
\fi
\bibpunct{(}{)}{,}{a}{}{,}
\makeindex
\citeindextrue

\newcommand{\getlength}[1]{\ifx#1\end \let\next=\relax
            \else\advance\count255 by1 \let\next=\getlength\fi \next}
\newcount\switch
\newcommand{\Endmat}{\ifnum\switch=0$\fi}
\newcommand{\ifnularg}[1]{ \count255=0 \getlength#1\end \ifnum\count255=0 }
\newcommand{\ifm}{\makebox{}\ifmmode}

\long\def\ifundefined#1#2#3{\expandafter\ifx\csname
  #1\endcsname\relax#2\else#3\fi}
\newcommand{\beq}   { \begin{eqnarray} }
\newcommand{\eeq}[1]{ \ifnularg{#1} end{eanarray} \else
                      \label{#1}\end{eqnarray}    \fi }
\newcommand{\hm}{\hphantom{-}}
\ifdraft
\newcommand{\Number}[1]{\ifnum#1<10\relax0\number#1\else\number#1\fi}
\newcommand{\isodate}{
\count151=\time
\divide\count151 by 60
\count151=\count151
\multiply\count151 by 60
\count152=\time
\advance\count152 by -\count151
\divide\count151 by 60
\count152=\count151
\multiply\count151 by 60
\count153=\time
\advance\count153 by -\count151
\Number{\year}.\Number{\month}.\Number{\day}--\Number{\count152}:\Number{\count153}
}
\fi
\begin{document}

\title{The Sixth VLBA Calibrator Survey --- VCS6}

\author{L. Petrov}
\affil{NVI, Inc., 7257 Hanover Pkw., Suite D, Greenbelt, MD 20770, USA}
\email{Leonid.Petrov@lpetrov.net}
\author{Y. Y. Kovalev}
\affil{Max-Plank-Institut f\"ur Radioastronomie,
       auf dem H\"ugel 69, D-53121 Bonn, Germany}
\affil{Astro Space Center of Lebedev Physical Institute,
       Profsoyuznaya 84/32, 117997 Moscow, Russia}
\email{ykovalev@mpifr-bonn.mpg.de}
\author{E. B. Fomalont}
\affil{National Radio Astronomy Observatory,
       520 Edgemont Road, Charlottesville, VA~22903--2475, USA}
\email{efomalon@nrao.edu}
\author{D. Gordon}
\affil{NVI, Inc., 7257 Hanover Pkw., Suite D, Greenbelt, MD 20770, USA}
\email{dgg@leo.gsfc.nasa.gov}

\begin{abstract}

\ifdraft
\par\hm\par \vspace{-13.5ex} \fbox{\Large\sf Draft of \isodate } \par\hm\par \vspace{4ex} 
\fi

  This paper presents the sixth part to the Very Long Baseline Array
(VLBA) Calibrator Survey. It contains the positions and maps of 264 sources 
of which 169 were not previously observed with very long baseline 
interferometry (VLBI). This survey, based on two 24 hour VLBA observing 
sessions, was focused on 1)~improving positions of 95 sources from previous 
VLBA Calibrator surveys that were observed either with very large a~priori 
position errors or were observed not long enough to get reliable positions 
and 2)~observing remaining new flat-spectrum sources with predicted 
correlated flux density in the range 100--200~mJy that were not observed 
in previous surveys. Source positions were derived from astrometric analysis 
of group delays determined at the 2.3 and 8.6~GHz frequency bands using the 
Calc/Solve software package. The VCS6 catalogue of source positions, plots 
of correlated flux density versus projected baseline length, contour plots 
and fits files of naturally weighted CLEAN images, as well as calibrated 
visibility function files are available on the Web at 
\url{http://vlbi.gsfc.nasa.gov/vcs6}.

\end{abstract}


\section{Introduction}
\label{s:introduction}

    This work is a continuation of the project of surveying the sky for
bright compact radio sources. These sources can be used as phase referencing
calibrators for imaging of weak objects with very long baseline interferometry
(VLBI) and as targets for space navigation, monitoring the Earth's rotation,
differential astrometry, and space geodesy. Precise positions of radio sources 
are needed for these applications. Since 1979 more than 4400 24~hour VLBI 
experiments have been scheduled in geodetic mode. These observations also 
determined parameters associated with the Earth orientation and rotation, 
antenna position and motions and other astrometric/geodetic parameters.  
By November 01, 2007, 1137 sources were observed in these experiments and 
1045 of them were detected. The observations are described by \citet{icrf98} 
and the latest catalog of 776 sources is given as the ICRF-Ext2 catalog 
\citep{icrf-ext2-2004}.

Because the astrometric and image quality of the radio sources (targets) 
improve with decreasing angular separation of the calibrator and the 
target, a higher density of sources than that from the ICRF catalogue was 
needed. Since 1994, 22 dedicated 24-hr experiments with the VLBA, called the 
VLBA calibrator survey (VCS), were made in order to increase the density 
of suitable VLBI phase reference sources \citep{vcs1,vcs2,vcs3,vcs4,vcs5}. 
In these experiments, 3601 sources with declinations $> -40^\circ$ were 
observed, including 764 objects previously observed under geodetic programs, 
and 3301 of them were detected. The catalogue of source positions derived 
from a single least square (LSQ) solution using all geodetic and VCS 
observations forms the pool of sources with positions at a milliarcsec
level of accuracy that is widely used for phase referencing observations,
for statistical analysis, and for other applications. Improving the precision
of this catalogue and an increase in the number of objects is an important
task.

  In this paper we present an extension of the VCS catalogue, called the
VCS6 catalogue. The objectives of this campaign were to improve the catalogue
of compact radio sources. Our approach to this problem was 
1)~to improve positions of sources observed in the previous VCS campaigns that 
had a)~poor a~priori positions; b)~were not observed long enough in the first 
two VCS1 24-hour experiments recorded at the 64~Mbit/sec rate; 
and 2)~to get positions of new sources. The new sources were taken from 
the lists of a)~intra-day variable sources observed in the framework of the 
MASIV survey \citep{masiv-2003}, not observed before with the VCS, and 
b)~leftovers from the list of candidate sources prepared for the VCS5 
campaign \citep{vcs5} that were not observed due to lack of resources.

  Since the observations, calibrations, astrometric solutions and imaging
are similar to that of VCS1--5, most of the details are described by
\cite{vcs1}, \cite{vcs3}, and \cite{vcs4}. In \S\ref{s:selection} we
describe the strategy for selecting 285 candidate sources observed in
two 24~hour sessions with the VLBA. In \S\ref{s:obs_anal} we briefly outline 
the observations and data processing. We present the VCS6 catalogue in 
\S\ref{s:catalogue}, and summarize our results in \S\ref{s:summary}.

\section{Source selection}
\label{s:selection} 

  Improvement of the astrometric catalogue of compact radio sources
can be made in two ways: 1)~improve of position accuracy of the sources with
coordinate errors significantly worse than the median error; 2)~increase 
the total number of sources. We selected both ways.

  It should be noted that in general the problem of improving the catalogue 
is mainly the question of optimization of resource allocation. Had we had 
unlimited resources, it would be sufficient to observe all sources for 
a long time and with the highest data rate the current technology can offer. 
But we would like to achieve the best catalogue improvement with the 
relatively small resources that were available.

  The formal position uncertainty of a source detected in two scans at all 
VLBA baselines is below 0.3~mas, which is believed to be the level of 
systematic errors (refer to \citealt*{icrf98} for justification of this limit) 
caused by mismodeling troposphere path delay and unaccounted source structure. 
At the same time, the formal uncertainties of 13\% of the sources in the 
astrometric pool are worse than 5~mas. This means there were not enough 
detections of these sources, especially at long baselines, for deriving 
precise coordinates. The longer the baseline, the more sensitive observations 
are to source position with a given signal to noise ratio. There may be 
a number of reasons why a source is not detected at long baselines. First, 
the source may be too resolved, and its correlated density at long baselines 
falls below the detection limit. This is the most common reason. Since for 
many sources the correlated flux density drops rapidly with increasing 
baseline length, this would require a significant improvement in sensitivity 
in order to detect the source at all baselines. A second reason is that the 
a~priori source positions were so bad that interferometric fringes were not 
found in the search window. A third reason is that the source was not 
scheduled at long baselines at all. This often happens when a scheduling 
algorithm for geodetic VLBI observations is used. Geodetic schedules are 
optimized to have uniform source sky coverage at each station every 1--3 
hours and this goal often contradict with having observations at long baseline 
for every targeted source. A fourth reason is instrumental failure or 
especially bad weather at one or more stations. 

  We selected sources from several lists. The first list marked ``r'' contains
68 sources that were observed in the first two 24-hour VCS1 experiments and 
had from 1 to 9 detections only. The VCS experiments in 1994--1995 were 
recorded at the 64~Mbit~s${}^{-1}$  rate, half the rate of later experiments, 
and each source was observed for 90~s. We re-observed them in two scans of 
188~s with 128~Mbit~s${}^{-1}$, i.e. lowered the detection limit by a factor 
of~2. The second list marked ``a'' contains 27 sources correlated with 
a~priori position errors ranging from $30''$ to $7'.5$. With such huge errors 
in a~priori positions, the process of fringe search may fail, especially at 
long baselines, because correction to the group delay and delay rate may be 
outside the search window. Even if fringes are detected, the fringe amplitude
and phase may be significantly distorted. The sources from lists ``r'' and 
``a'' have a good chance to be detected at long baselines, therefore they
are considered to be the best candidates for improvement with a minimal
resource allocation.

  The second way to improve an astrometric catalogue is to observe new 
sources. There is a strong evidence \citep[e.g.,][]{jm2001,r2001,dtb2002} 
that  the intra-day variability (IDV) is caused by scattering in the 
galactic media. \citet{ojha-idv} observed with the VLBA a sample of 75 IDV 
sources with flux density below 300~mJy and made images; they confirmed that 
these sources have a very compact, core-dominated morphology. 
\citet{2cmPaperIV} analyzed properties of 250 strong active galactic nuclei 
including 43 known IDV sources; they have shown that the VLBI compactness 
parameter reaches its highest value for the IDV sub-sample of objects. 
Unfortunately, \citet{ojha-idv} did not make images, even the source list, 
publicly available in digital form, so we were unable to use their results 
for source selection. H.E.~Bignall provided us preliminary estimates of the 
mean flux density at 5~GHz from VLA observations of the final sample of 
482 sources (private communication, 2006) observed in the MASIV program 
\citep{masiv-2003}. We removed from this list the sources previously \
observed with the VLBA in astrometry/geodesy mode and ordered  the 
remaining 124 sources by decreasing their flux densities. The first  
top 30~objects with the highest flux densities derived from analysis of 
the MASIV VLA observations exceeding 130~mJy at 5~GHz were scheduled in the
VCS6 campaign in two scans, each with integration time of 193~s. The
list of these sources is marked as ``i''.

  The other list of candidate sources are the leftover sources from the VCS5 
campaign which aimed to observe all the remaining flat-spectrum sources with 
expected correlated flux density $> 200$~mJy \citep{vcs5}. For each source 
the probability of detection was computed based on published multi-frequency
single dish flux densities. The probability of detection at the VLBA was 
computed according to the algorithm proposed by \citet{vcs4} that is based 
on estimates of the extrapolated flux density at the frequency 8.6~GHz and 
the spectral index. Unfortunately, these estimates are often either 
incomplete or unreliable, especially for sources with flux density below 
300~mJy. We visually inspected each spectrum plot and assigned to each 
spectrum a class in accordance to our estimation of a probability to detect 
correlated flux density greater than 200~mJy at 8.6~GHz. All the sources 
from the first class with the highest probability of detection were observed 
by \citet{vcs5}. In this paper we have observed 160~objects of the second 
class with an intermediate probability (list ``v'').

  In addition to target sources, tropospheric calibrators have to be 
observed extensively in astrometric experiments. The list of troposphere
calibrators\footnote{\url{http://vlbi.gsfc.nasa.gov/vcs/tropo\_cal.html}}
was selected from the sources which, according to the 2~cm VLBA survey
results \citep{2cmPaperIV}, showed the greatest compactness index, i.e.
the ratio of the correlated flux density measured at long VLBA spacings
to the flux density integrated over the VLBA image. 

\section{Observations and data analysis}
\label{s:obs_anal} 

  Observations were performed in two 24~hour sessions with the VLBA 
on 2006 December 18 and 2007 January 11. The target sources were observed 
in a sequence designed to minimize loss of time from antenna slewing. 
In addition to these objects, observations of 3--4 strong sources from 
a list of 62 tropospheric calibrators
were made every 1--1.5~hours during the sessions. These observations were 
scheduled in such a way, that at each VLBA station at least one of these 
sources was observed at an elevation angle less than 20\degr, and at least 
one at an elevation angle greater than 50\degr. The purpose of observing 
tropospheric calibrators is twofold. First, they  significantly  improve 
separation of variables that parameterize variable zenith path delay modeled 
as an expansion over the B-spline basis with equidistant nodes with a time 
span of 20~minutes. Second, their positions are listed in the ICRF 
catalogue \citep{icrf98}, so they tie positions of VCS6 sources to this 
catalogue. In total, 347 sources were observed, including 285 targeted 
objects. The antennas were scheduled to be on-source 70.2\% of the time.

  Similar to the previous VLBA Calibrator Survey observing campaigns,
we used the VLBA dual-frequency geodetic mode, observing simultaneously at 
2.3~GHz and 8.6~GHz, right circular polarization. Each band was separated 
into four 8~MHz channels (IFs) which spanned 140~MHz at 2.3~GHz and 490~MHz 
at 8.6~GHz (Table \ref{t:frq}), in order to provide precise measurements 
of group delays for astrometric processing. The data were correlated with 
an accumulation period of 0.5~second in 64~frequency channels per IF 
in order to provide a wide window for fringe searche.

\begin{deluxetable}{cc}
\tablecaption{The range of intermediate frequencies
\label{t:frq}
}
\tablewidth{-106.51149pt}
\tablecolumns{2}
\tablehead{
\colhead{IF} & \colhead{Frequency range (MHz)} \\
\colhead{(1)} & \colhead{(2)}
}
\startdata
1 & 2232.99 --- 2240.99   \\
2 & 2262.99 --- 2270.99   \\
3 & 2352.99 --- 2360.99   \\
4 & 2372.99 --- 2380.99   \\
5 & 8405.99 --- 8413.99   \\
6 & 8475.99 --- 8483.99   \\
7 & 8790.99 --- 8898.99   \\
8 & 8895.99 --- 8903.99  
\enddata
\end{deluxetable}

\begin{deluxetable}{crr}
\tablecaption{Statistics of detection of the sources observed in the VCS6
campaign \label{t:t1}
}
\tablewidth{-135.52678pt}
\tablecolumns{3}
\tablehead{
\colhead{Source list} & \colhead{Detected} & \colhead{Non-detected}\\
\colhead{(1)} & \colhead{(2)} & \colhead{(3)}
}
\startdata
a &   27  &   0  \\
r &   68  &   0  \\
v &  139  &  21  \\
i &   30  &   0  \\
T &   62  &   0 
\enddata
\tablecomments{
Column designation:
(1) The name of the list;
(2) the number of detected sources;
(3) the number of not detected sources.
}
\end{deluxetable}

  Processing of the VLBA correlator output was done in three steps. In the
first step the data were calibrated and fringed using the Astronomical
Image Processing System (AIPS) \citep{aips} in a standard way. An
amplitude shift was present in the calibrated data because of the
following. Our IFs are widely spread over receiver bands
(Table~\ref{t:frq}) while the VLBA S~band and X~band gain-curve
parameters applied are measured around 2275~GHz and 8425~MHz
respectively \citep{VLBA_summ}, and the noise diode spectrum is not
ideally flat. We have used strong flat-spectrum sources in the sample in
order to estimate global amplitude correction factors which differ from
1.00 by more than 0.09 (Table~\ref{t:amp_cor}). These coefficients were
applied to the data in both observing sessions.

  In the second step, data were imported to the Caltech DIFMAP 
package~\citep{difmap}, $uv$ data flagged, and images were produced using 
automated hybrid imaging procedure originally suggested by Greg 
Taylor~\citep{difmap-script} which we optimized for our experiment. We were 
able to reach the VLBA image thermal noise level for most of our CLEAN images
\citep{VLBA_summ}. Errors on our estimates of correlated flux density
values for sources stronger than $\sim$100~mJy were dominated by the
accuracy of amplitude calibration, which for the VLBA, according to
\citet{VLBA_summ}, is at the level of 5\,\% at 1~GHz to 10~GHz.
Our error estimate was confirmed by comparison of the flux densities
integrated over the VLBA images with the single-dish flux densities
which we measured with \mbox{RATAN--600} in 2006 and 2007 for slowly
varying sources without extended structure. The methods of RATAN--600
single-dish observations and data processing can be found in 
\citet{Kovalev_etal99}.

\begin{deluxetable}{lcc}
\tablecaption{Amplitude correction coefficients
\label{t:amp_cor}
}
\tablewidth{-110.09534pt}
\tablecolumns{3}
\tablehead{
\colhead{Antena} & \colhead{IF} & \colhead{Coefficient}\\
\colhead{(1)} & \colhead{(2)} & \colhead{(3)}
}
\startdata
Bruster     & 4 & 1.75 \\
Los~Alamos  & 3 & 0.83 \\
Los~Alamos  & 4 & 0.70 \\
Ovens Valey & 4 & 0.90 
\enddata
\end{deluxetable}

\ifdraft
\begin{figure}[p]
\else
\begin{figure}[b]
\fi
  \par\hphantom{a}\par\vspace{3ex}\par
   \begin{center}
      \resizebox{0.95\hsize}{!}{\includegraphics[trim=0cm 0cm 0cm 0cm,width=0.8\textwidth]{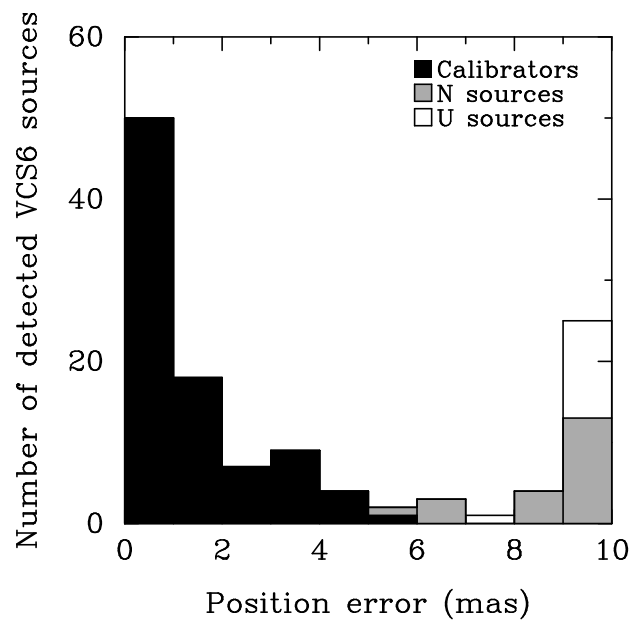}}
   \end{center}
   \caption{Histogram of semi-major error ellipse of position errors.
            The last bin shows errors exceeding 9~mas.
            See explanation of different assigned source classes in
            \S~\ref{s:catalogue}.}
   \label{f:errhist}
  \par\hphantom{a}\par\vspace{-4ex}\par
\end{figure}

\ifpre
  \begin{deluxetable*}{ l @{\hspace{0.2em}} l l l r r r r @{\hspace{0.2em}} r r r r r l l}[h]
\else
  \begin{deluxetable}{ l l l l r r r r r r r r r l l}
\fi
\tablecaption{The VCS6 catalogue \label{t:cat}}
\tablewidth{0pt}
\tablecolumns{15}
\ifpre \relax \else \rotate \fi
\tablehead{
   \colhead{} &
   \multicolumn{2}{c}{Source name}                   &
   \multicolumn{2}{c}{J2000.0 Coordinates}           &
   \multicolumn{3}{c}{Errors (mas)}                  &
   \colhead{}                                        &
   \multicolumn{4}{c}{Correlated flux density (Jy)}  &
   \colhead{}
   \vspace{0.5ex} \\
   \multicolumn{9}{c}{}         &
   \multicolumn{2}{c}{8.6 GHz}  &
   \multicolumn{2}{c}{2.3 GHz}  \\
   \colhead{Class}    &
   \colhead{IVS}      &
   \colhead{IAU}      &
   \colhead{Right ascension} &
   \colhead{Declination}     &
   \colhead{$\Delta \alpha$} &
   \colhead{$\Delta \delta$} &
   \colhead{Corr}   &
   \colhead{\# Obs} &
   \colhead{Total } &
   \colhead{Unres } &
   \colhead{Total } &
   \colhead{Unres } &
   \colhead{Band}   &
   \colhead{List}
   \vspace{0.5ex} \\
   \colhead{(1)}    &
   \colhead{(2)}    &
   \colhead{(3)}    &
   \colhead{(4)}    &
   \colhead{(5)}    &
   \colhead{(6)}    &
   \colhead{(7)}    &
   \colhead{(8)}    &
   \colhead{\hfill (9)}    &
   \colhead{(10)}   &
   \colhead{(11)}   &
   \colhead{(12)}   &
   \colhead{(13)}   &
   \colhead{(14)}   &
   \colhead{(15)}
   }
\startdata
N  & 0000$-$199 & J0003$-$1941 & 00 03 15.949411 & $-$19 41 50.40190 &  2.59  &  6.88  & -0.814  &    11  & $ 0.12 $ & $<0.06 $ & $  0.20 $ & $ 0.14 $ & X/S & v \vspace{0.5ex} \\
C  & 0003$+$123 & J0006$+$1235 & 00 06 23.056110 & $+$12 35 53.09745 &  0.92  &  1.08  &  0.280  &    24  & $ 0.16 $ & $<0.06 $ & $  0.13 $ & $<0.06 $ & X/S & v \vspace{0.5ex} \\
C  & 0005$+$683 & J0008$+$6837 & 00 08 33.472899 & $+$68 37 22.04848 &  1.81  &  0.58  & -0.190  &    90  & $ 0.43 $ & $<0.06 $ & $  0.28 $ & \nodata  & X   & r \vspace{0.5ex} \\
C  & 0007$+$439 & J0010$+$4412 & 00 10 30.046481 & $+$44 12 42.50407 &  0.52  &  0.97  &  0.295  &    48  & $ 0.16 $ & $ 0.07 $ & $  0.19 $ & $ 0.14 $ & X/S & i \vspace{0.5ex} \\
C  & 0008$-$300 & J0010$-$2945 & 00 10 45.177362 & $-$29 45 13.17767 &  0.68  &  2.25  & -0.352  &    43  & $ 0.26 $ & $ 0.09 $ & $  0.56 $ & $ 0.49 $ & X/S & v \vspace{0.5ex} \\
C  & 0009$-$148 & J0011$-$1434 & 00 11 40.455912 & $-$14 34 04.63437 &  0.77  &  1.28  &  0.356  &    38  & $ 0.15 $ & $<0.06 $ & $  0.20 $ & $ 0.14 $ & X/S & v \vspace{0.5ex} \\
C  & 0010$-$155 & J0013$-$1513 & 00 13 20.701815 & $-$15 13 47.78346 &  1.16  &  2.69  & -0.334  &    16  & $ 0.13 $ & $<0.06 $ & $  0.15 $ & $ 0.12 $ & X/S & v \vspace{0.5ex} \\
C  & 0013$-$184 & J0015$-$1807 & 00 15 34.324495 & $-$18 07 25.58298 &  1.63  &  3.08  & -0.608  &    20  & $ 0.15 $ & $<0.06 $ & $  0.32 $ & $ 0.15 $ & X/S & v \vspace{0.5ex} \\
U  & 0016$-$223 & J0019$-$2205 & 00 19 22.939314 & $-$22 05 19.75395 & 20.67  & 59.79  &  0.977  &     2  & $ 0.09 $ & $ 0.07 $ & $  0.11 $ & \nodata  & X   & v \vspace{0.5ex} \\
C  & 0018$+$729 & J0021$+$7312 & 00 21 27.374710 & $+$73 12 41.93114 &  6.68  &  1.48  &  0.251  &    24  & $ 0.22 $ & \nodata  & $  0.61 $ & $ 0.15 $ & X/S & r \vspace{0.5ex}
\enddata
\tablecomments{\rm
               Table~\ref{t:cat} is presented in its entirety in the electronic
               edition of the Astronomical Journal. A portion is shown here
               for guidance regarding its form and contents. Units of right
               ascension are hours, minutes and seconds, units of declination
               are degrees, minutes and seconds. Assigned source class in (1) 
               is `C' for calibrator, `N' for non-calibrator with reliable 
               coordinates, `U' for non-calibrator with unreliable coordinates; 
               see \S\,\ref{s:catalogue} for details.
              }
\ifpre
  \end{deluxetable*}
\else
  \end{deluxetable}
\fi

  In the third step, the data were imported to the Calc/Solve software,
group delay ambiguities were resolved, outliers eliminated and
coordinates of new sources were adjusted using ionosphere-free
combinations of X~band and S~band group delay observables of the two VCS6
sessions, 22 VCS1--5 experiments and 4453 twenty four hour International
VLBI Service for astrometry and geodesy (IVS)
experiments\footnote{\url{http://vlbi.gsfc.nasa.gov/astro}} in a single
least square solution. A boundary condition requiring zero net-rotation
of new coordinates of the 212 sources listed as ``defining'' in  the
ICRF catalogue with the respect to their positions from that catalogue
was imposed in order to select a unique solution of differential
equations of photon propagation.

  Among 285 observed sources, 264 were detected (refer to 
Table~\ref{t:t1}), including 169 new sources. Of the 95 sources 
re-observed in VCS6, all were detected, and the image and astrometric 
precision were significantly improved. Among these 95 re-observed 
objects, 27 sources had position accuracies better than 5~mas before the 
VCS6 campaign, after the VCS6 campaign this number increased to 63 
objects. The histogram of source position errors is presented in 
Figure~\ref{f:errhist}.

\section{VCS6 catalogue}
\label{s:catalogue}

  The VCS6 catalogue is listed in Table~\ref{t:cat}. Although the positions 
of 3913 sources were adjusted in the LSQ solution that used all observations
from 4400 observing sesions, only coordinates of targeted sources observed 
in the VCS6 campaign are presented. Some of targeted sources were alswo 
observed in previous campaigns. 

\begin{figure}[b!]
  \begin{center}
        \resizebox{1.0\hsize}{!}{\includegraphics[trim=-0.5cm 0cm 0.5cm 0cm,angle=270]{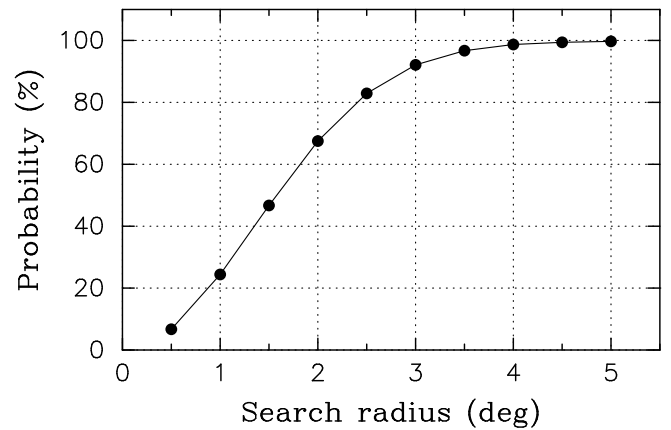}}
  \end{center}
  \figcaption{The probability (filled circles) of finding a calibrator 
              in any given direction within a circle of a given radius, 
              north of declination $-40\degr$. All 2853 sources from more 
              than 4400 IVS geodetic/astrometric sessions and 24 VCS1 
              to VCS6 VLBA sessions that are classified as calibrators 
              are taken into account.
              \label{f:prob}
             }
\end{figure}

\ifdraft
\begin{figure*}[p]
\else
\begin{figure*}[h]
\fi
\begin{center}
\resizebox{1.0\hsize}{!}
{
  \includegraphics[trim=-0.5cm 5cm 0cm 0cm,width=6.0cm]{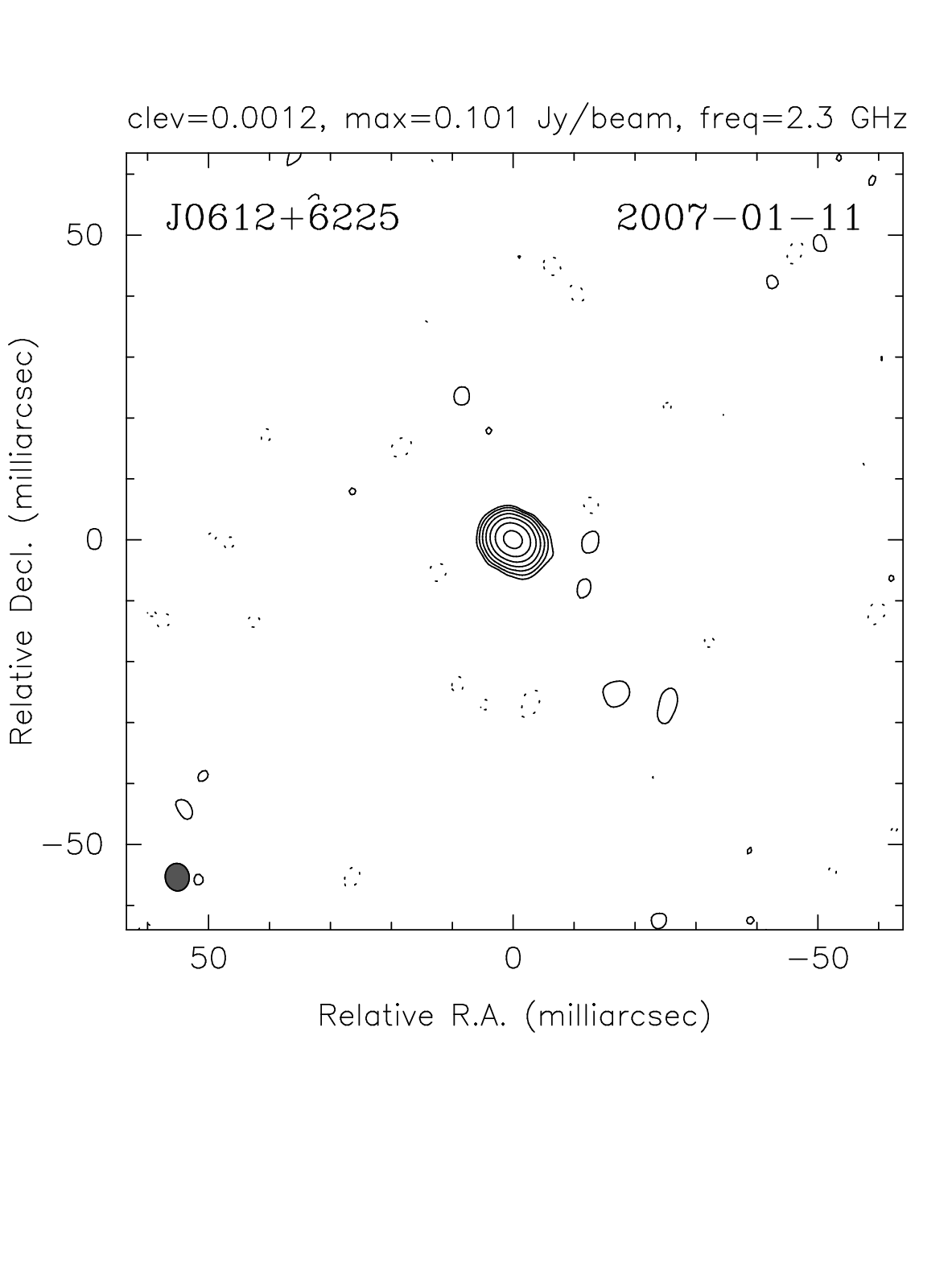}
  \includegraphics[trim=-1.0cm 5cm 0cm 0cm,width=6.0cm]{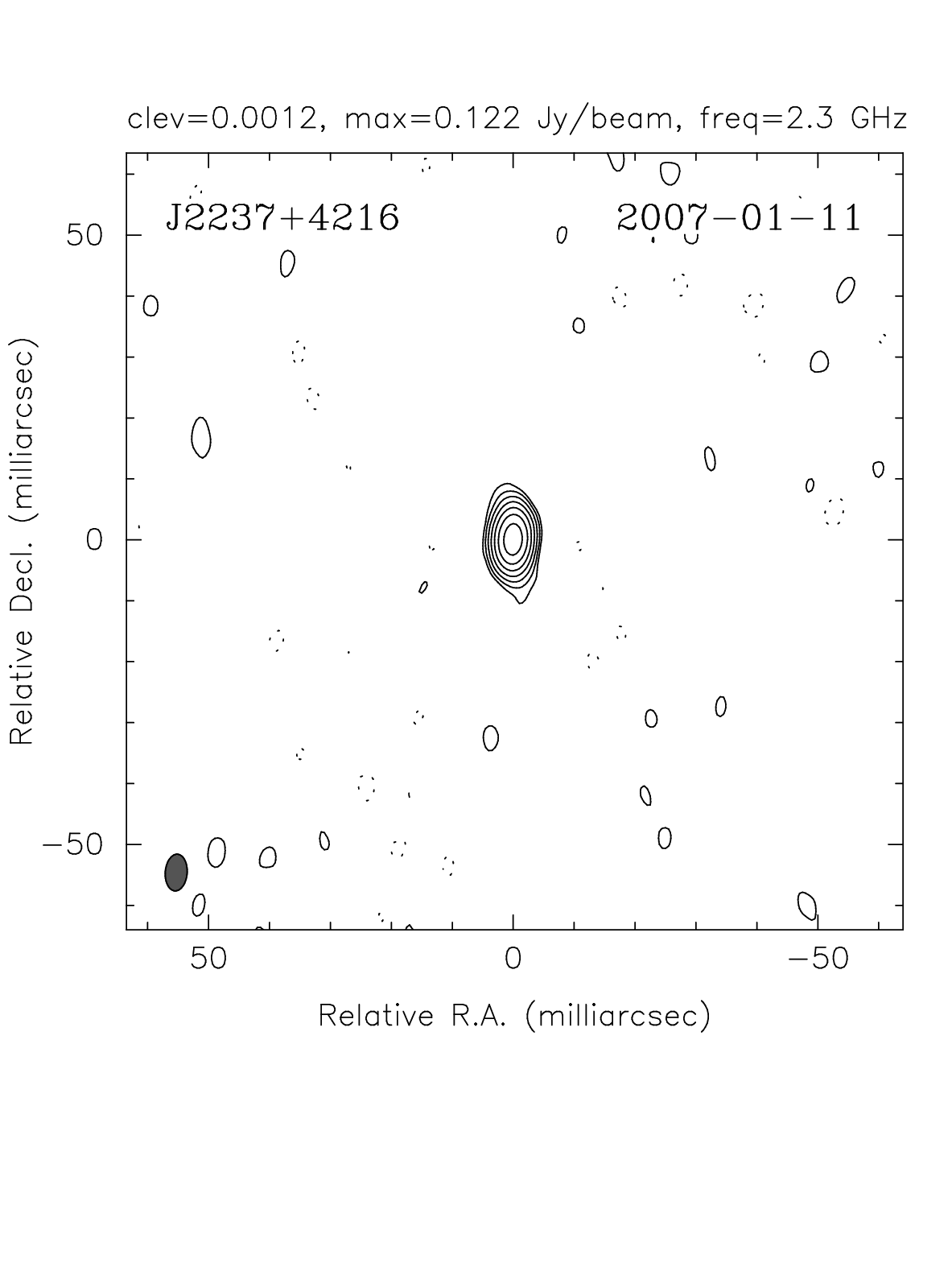}
  \includegraphics[trim=-1.0cm 5cm 0cm 0cm,width=6.0cm]{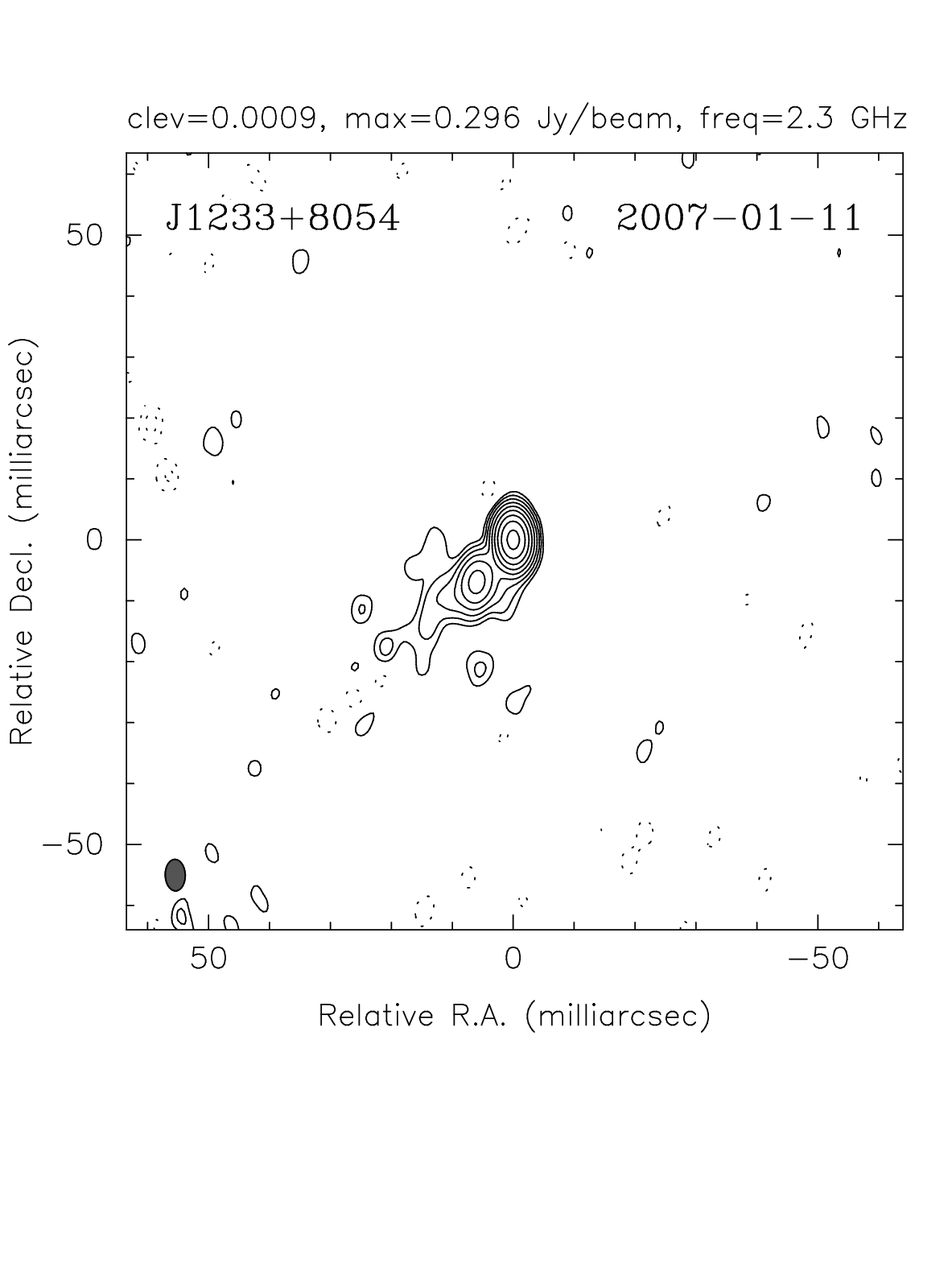}
}
\resizebox{1.0\hsize}{!}
{ 
  \includegraphics[trim=0cm  0.0cm 0cm 0cm,angle=270,width=6.0cm]{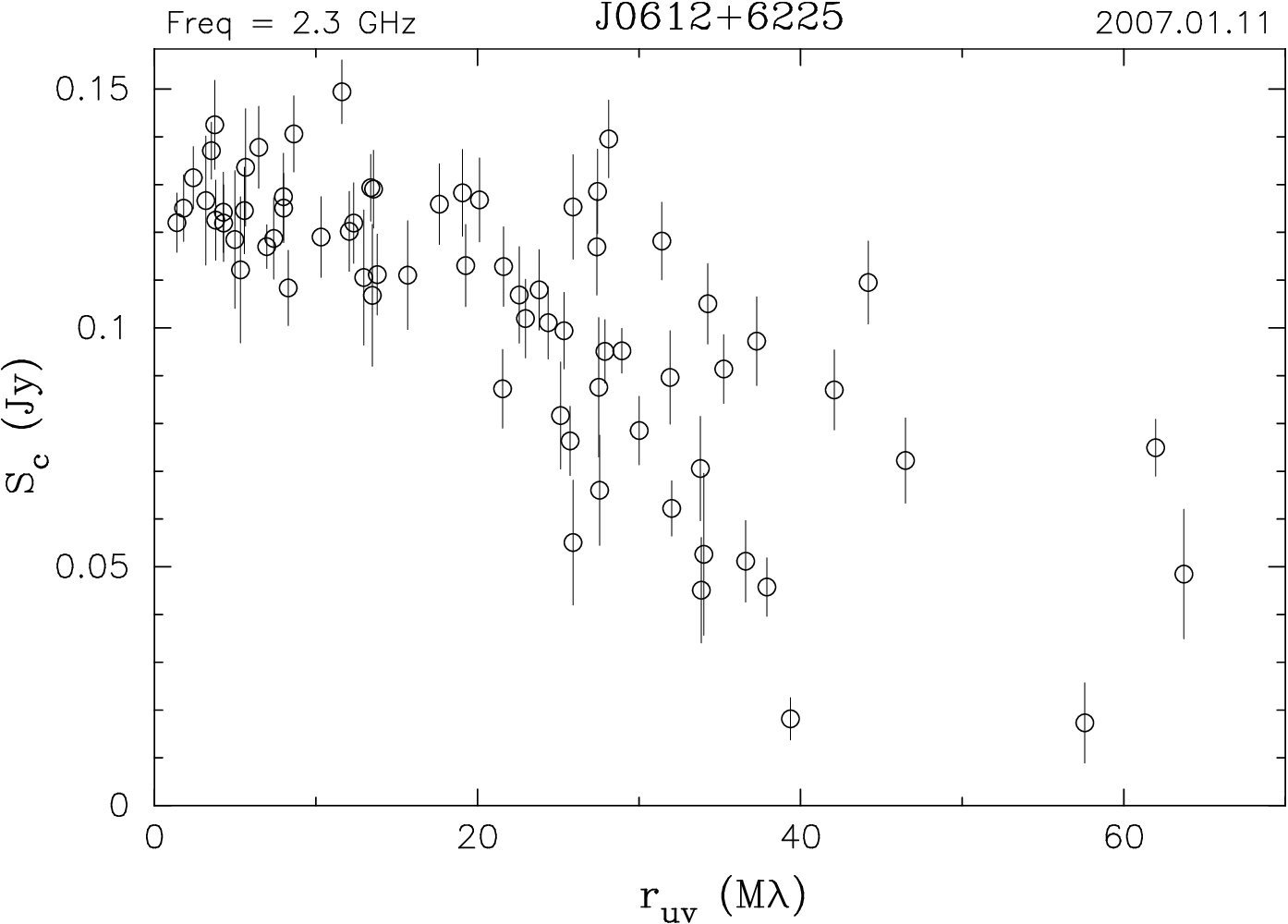}
  \includegraphics[trim=0cm -0.5cm 0cm 0cm,angle=270,width=6.0cm]{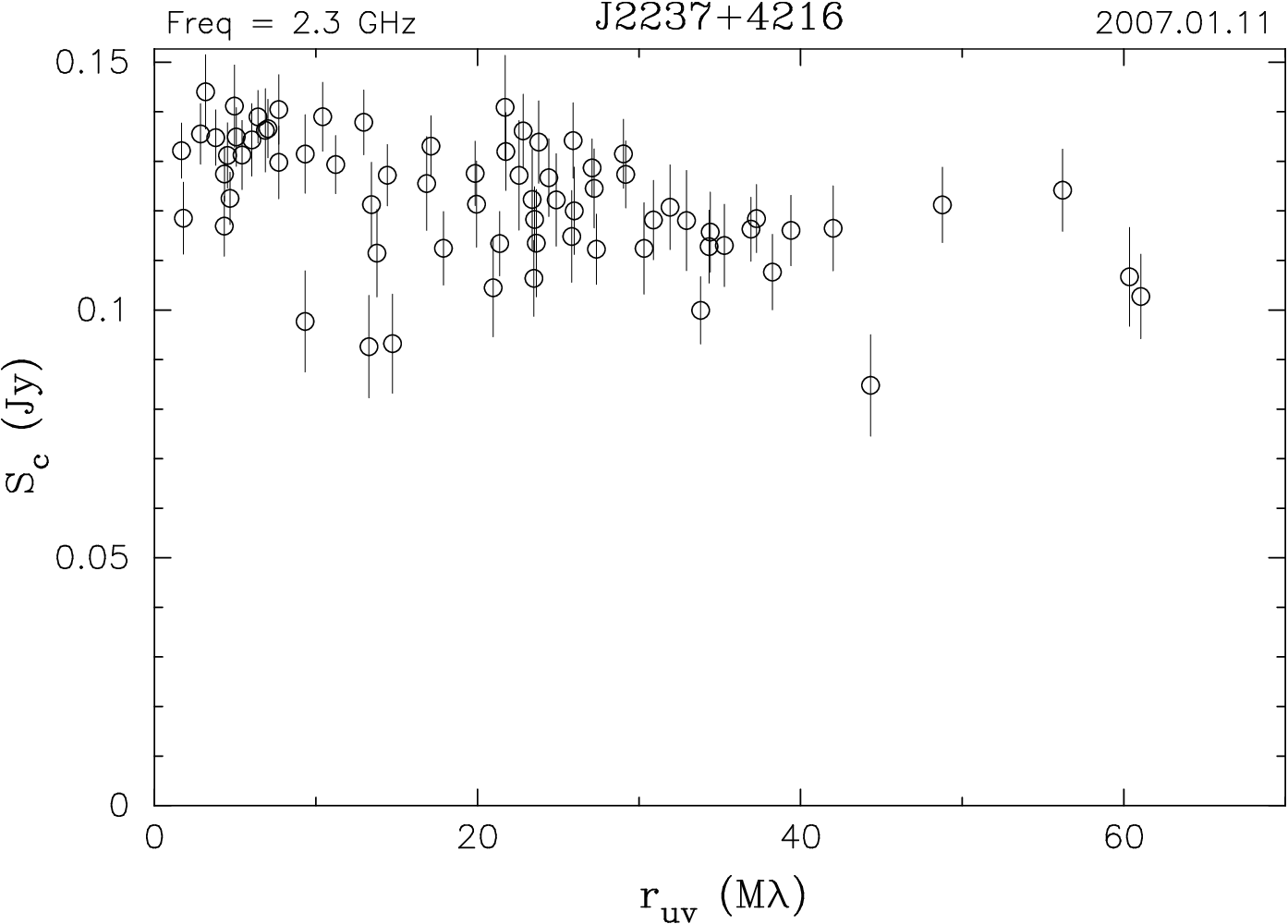}
  \includegraphics[trim=0cm -0.5cm 0cm 0cm,angle=270,width=6.0cm]{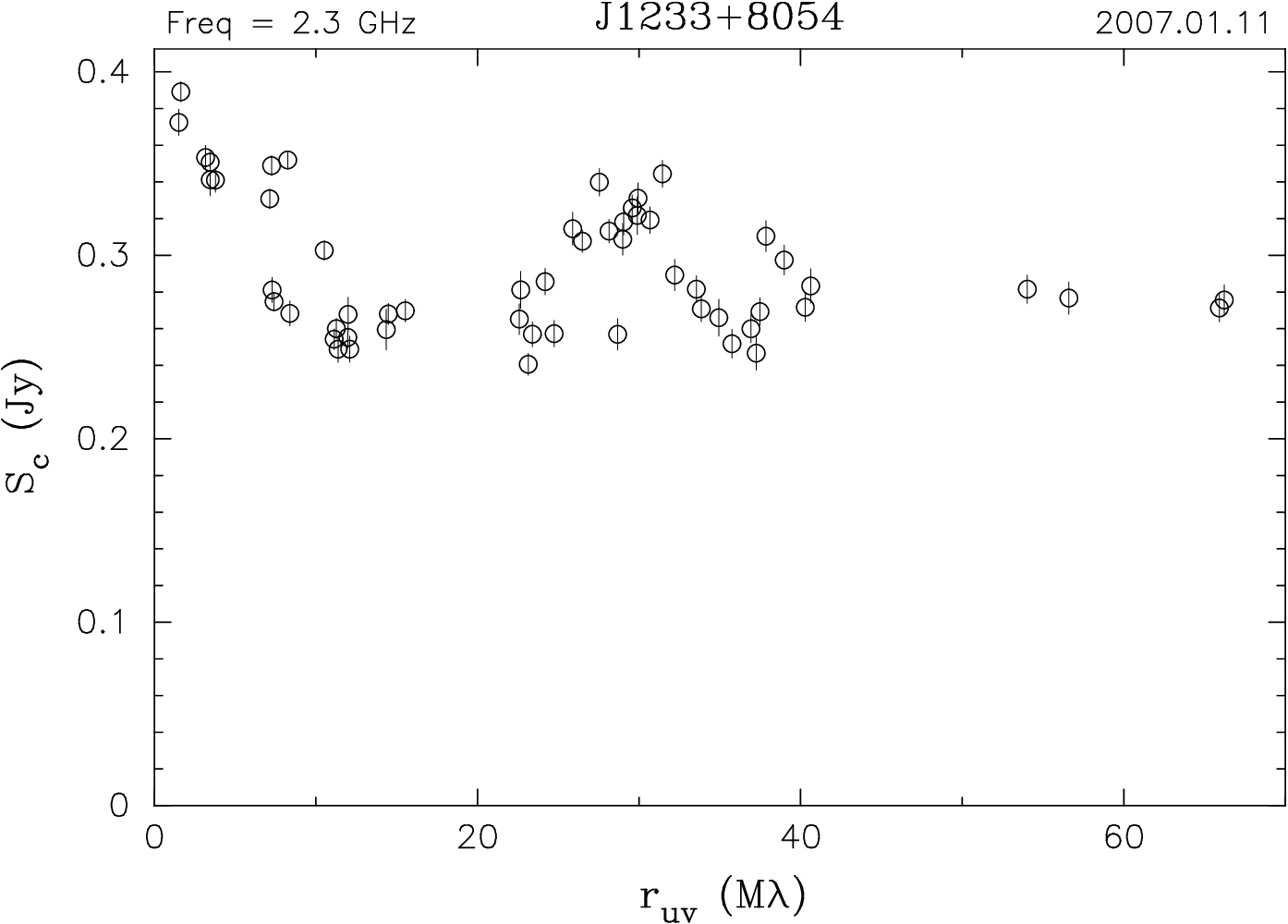}
}
\resizebox{1.0\hsize}{!}
{
  \includegraphics[trim=-0.5cm 5cm 0cm 1cm,clip,width=6.0cm]{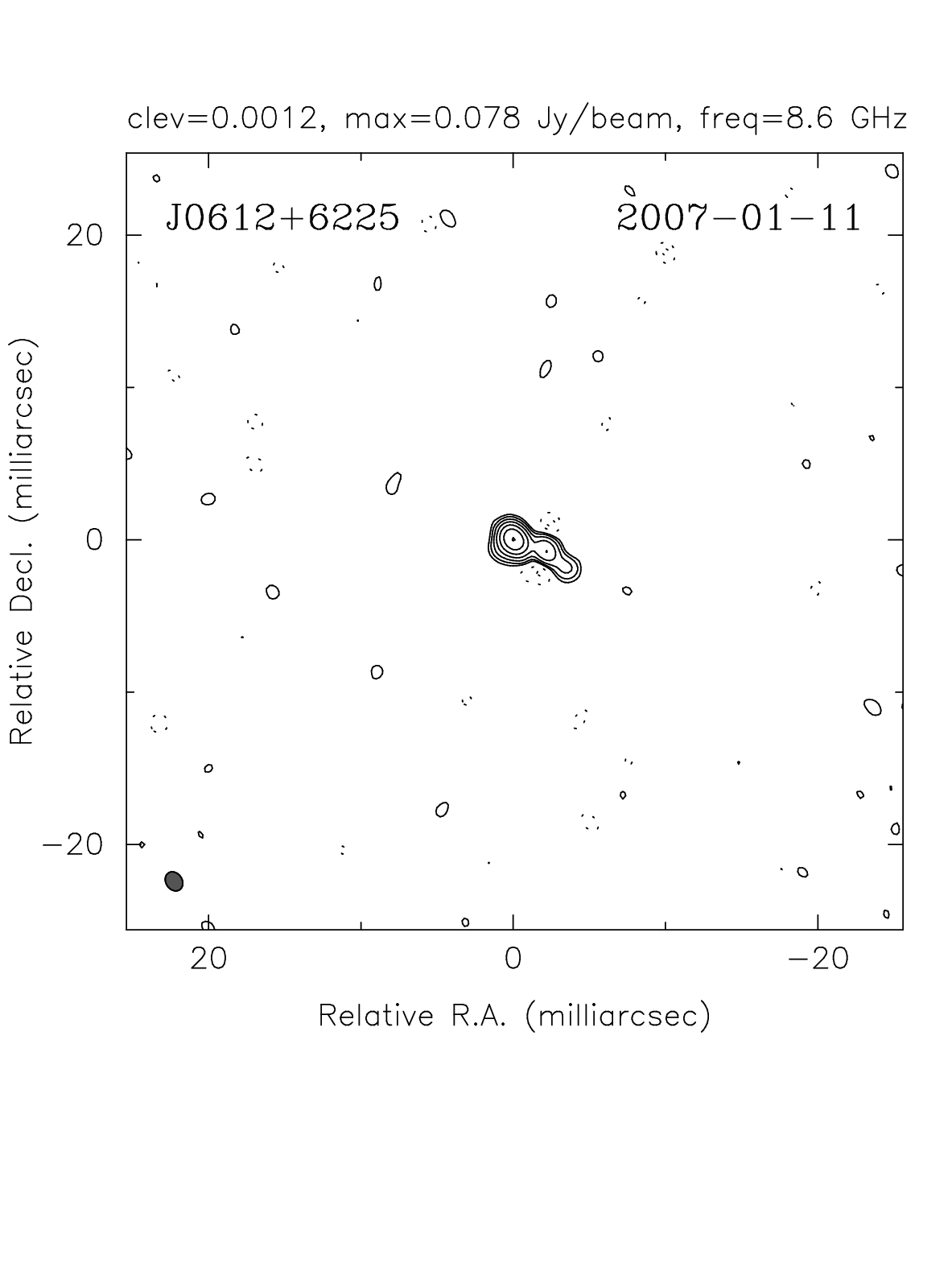}
  \includegraphics[trim=-1.0cm 5cm 0cm 1cm,clip,width=6.0cm]{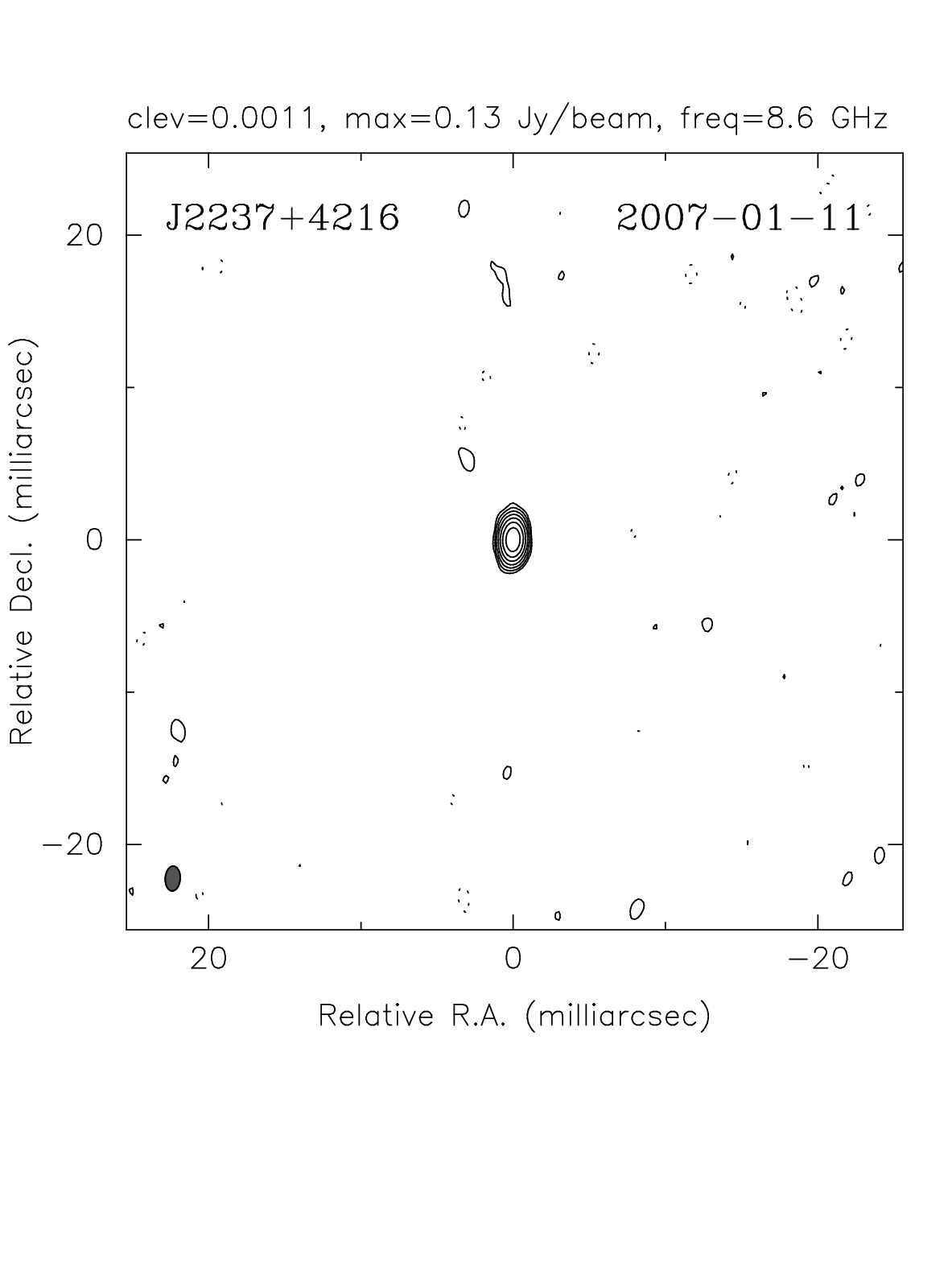}
  \includegraphics[trim=-1.0cm 5cm 0cm 1cm,clip,width=6.0cm]{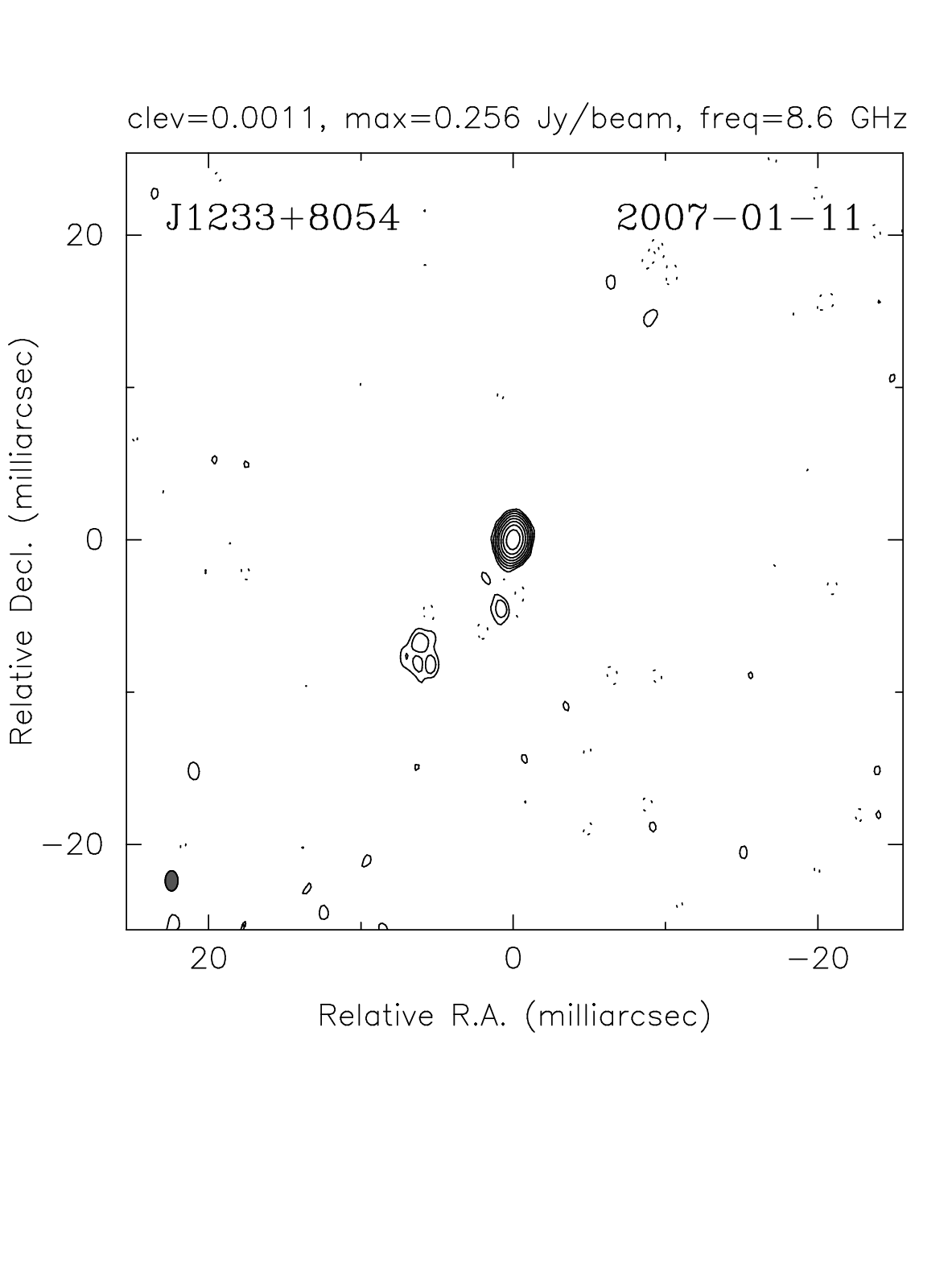}
}
\resizebox{1.0\hsize}{!}
{
   \includegraphics[trim=0cm  0.0cm 0cm 0cm,angle=270,width=6.0cm]{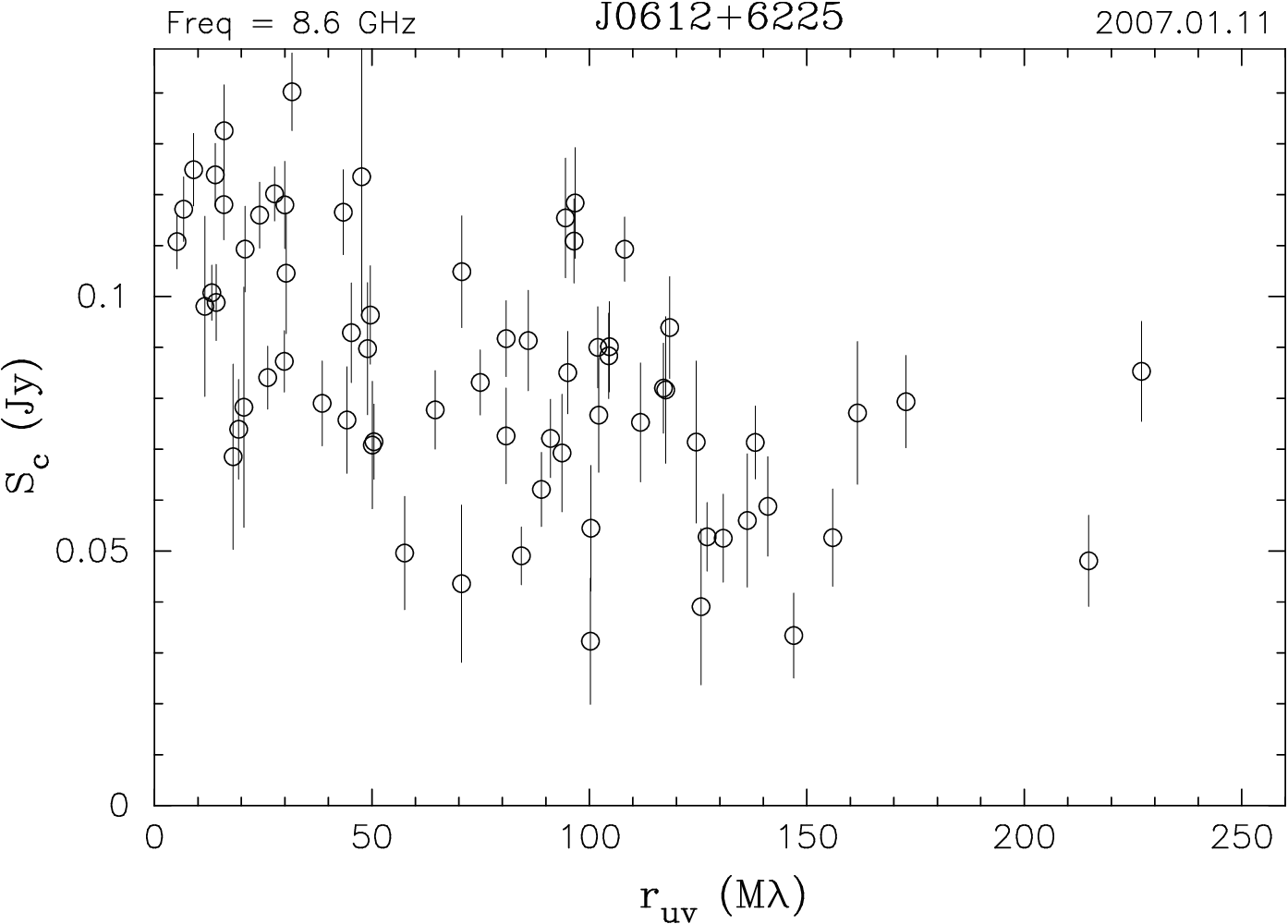}
   \includegraphics[trim=0cm -0.5cm 0cm 0cm,angle=270,width=6.0cm]{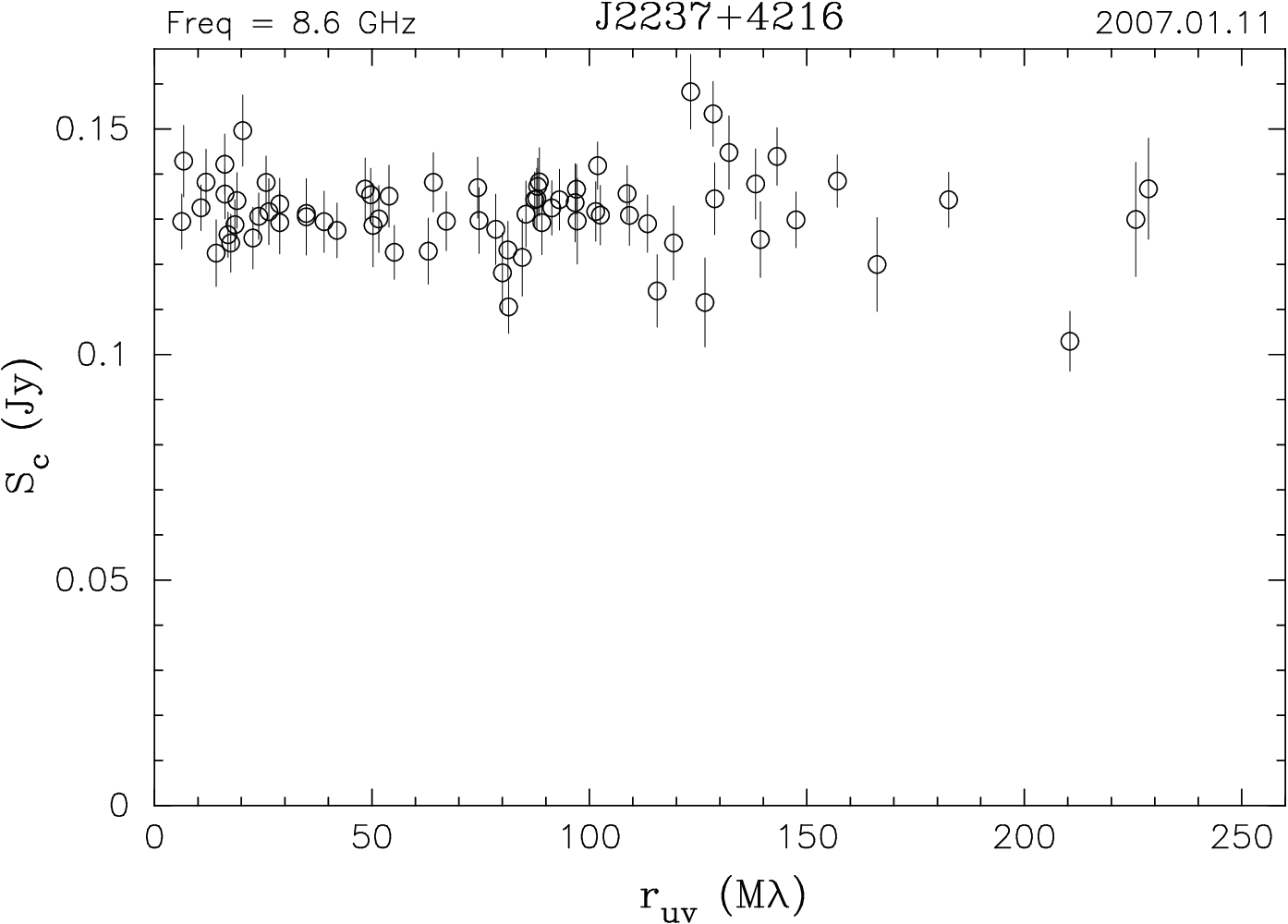}
   \includegraphics[trim=0cm -0.5cm 0cm 0cm,angle=270,width=6.0cm]{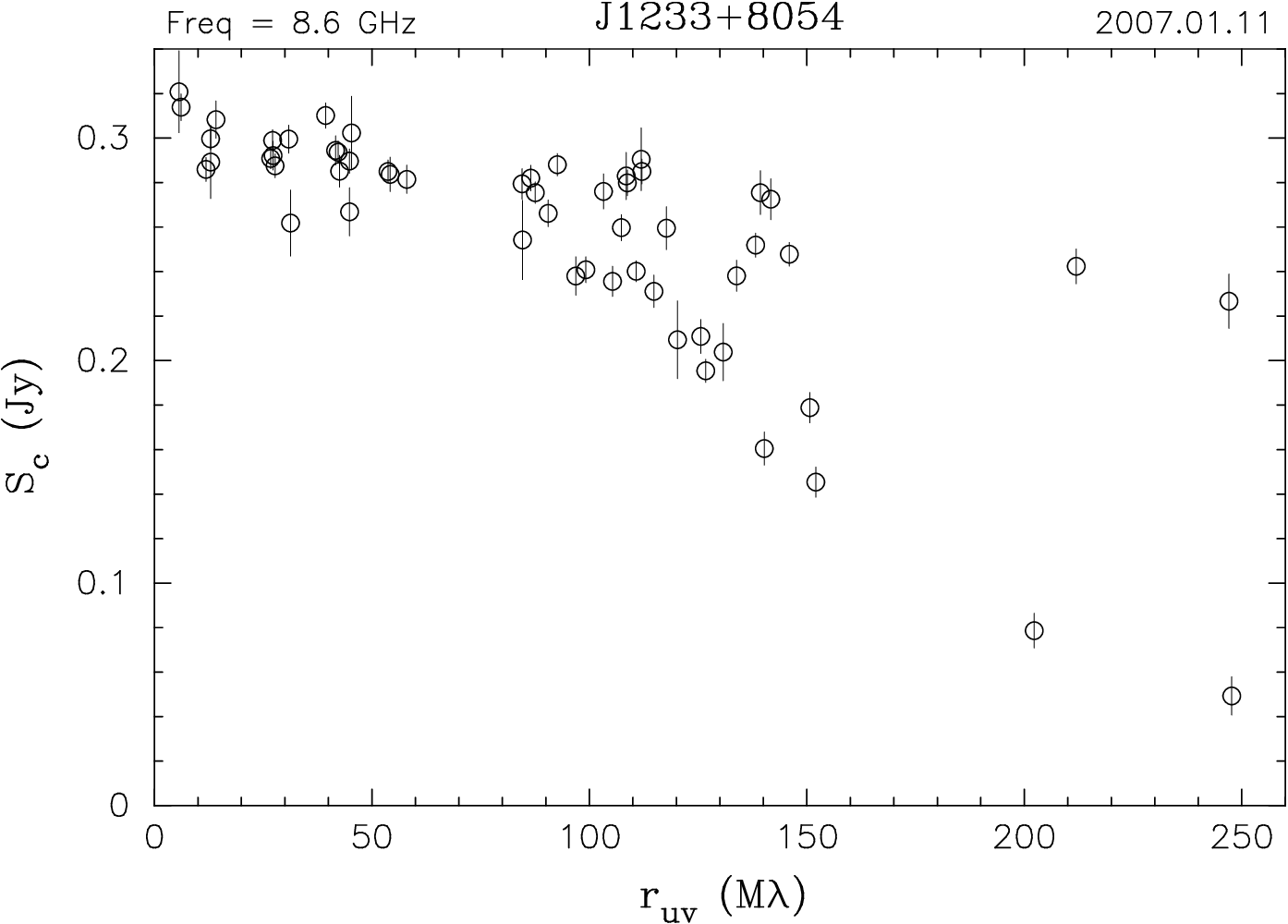}
}
\end{center}
\figcaption{\footnotesize 
             From top to bottom.
             {\em Row~1:}
             Naturally weighted CLEAN images at S~band (2.3~GHz). 
             The lowest contour is plotted at the level given by ``clev'' 
             in each panel title (Jy/beam), the peak brightness is given by 
             ``max'' (Jy/beam). The contour levels increase by factors of two. 
             The dashed contours indicate negative flux. The beam is shown 
             in the bottom left corner of the images. 
             {\em Row~2:}
             Dependence of the correlated flux density at S~band on projected
             spacing. Each point represents a coherent
             average over one 2~min observation on an individual interferometer
             baseline. The error bars represent only the statistical errors.
             {\em Row~3:} Naturally
             weighted CLEAN images at X~band (8.6~GHz).
             {\em Row~4:} Dependence of the
             correlated flux density at X~band on projected spacing.
             \label{f:images}
}
\end{figure*}

  The first column gives the calibrator class: ``C''~--- calibrator, i.e. 
the semi-major axis of the inflated error ellipse is less than 5 mas and more 
than 8 good pairs of X/S group delay measurements are available. If a source 
does not satisfy these criteria, it is assigned either class ``N'' --- 
non-calibrator with reliable positions, i.e. more than 8 good group delay 
measurements at X or S band are available; or class ``U'' --- non-calibrator 
with unreliable positions, i.e. less than 8 detections at any band are 
available, and therefore, there is a risk that group delay ambiguities were 
resolved  incorrectly. The second and third columns give IVS source name 
(B1950 notation), and IAU name (J2000 notation). The fourth and fifth columns 
give source coordinates at the J2000.0 epoch. Columns /6/ and /7/ give inflated
source position uncertainties in right ascension (without $\cos\delta$ factor) 
and declination in mas, and column /8/ gives the correlation coefficient 
between the errors in right ascension and declination. The number of group 
delays used for position determination is listed in column /9/. Columns 
/10/ and /12/ give the estimate of the flux density integrated over 
entire map in Janskies at X and S~band respectively. This estimate is computed 
as a sum of all CLEAN components if a CLEAN image was produced. If we did 
not have enough detections of the visibility function to produce a reliable 
image, the integrated flux density is estimated as the median of the correlated 
flux density measured at projected spacings less than 25 and 7~M$\lambda$ 
for X~and S~bands respectively. The integrated flux density is the source 
total flux density with spatial frequencies less than 4~M$\lambda$ at X~band 
and 1~M$\lambda$ at S~band filtered out, or in other words, this is the flux 
density from all details of a source with size less than 50~mas at X~band and 
200~mas at S~band. Column /11/ and /13/ give the flux density of unresolved 
components estimated as the median of correlated flux density values measured 
at projected spacings greater than 170~M$\lambda$ for X~band and greater than 
45~M$\lambda$ for S~band. For some sources no estimates of the integrated 
and/or unresolved flux density are presented, because either no data were 
collected at the baselines used in calculations, or these data were 
unreliable. Column /14/ gives the data type used for position estimation: 
X/S stands for ionosphere-free linear combination of X and S wide-band group 
delays; X stands for X~band only group delays; and S stands for S~band only 
group delays. Some sources which yielded less than 8 pairs of X and S band 
group delay observables had 2 or more observations at X and/or S band 
observations. For these sources either X-band or S-band only estimates of 
coordinates are listed in the VCS6 catalogue, whichever uncertainty is 
less. Column /15/ gives the name of the input source list.

Table~\ref{t:nondetected} presents 21 sources not detected in VCS6 VLBA
observations. Source positions used for observations and correlation are
presented. They are taken from the NVSS catalogue \citep{nvss}. The correlated 
flux density for the non-detected sources is estimated to be less than 
60~mJy at 2.3~GHz and 8.6~GHz.

\begin{deluxetable}{l l r r c}
\tablewidth{0pt}
\tablecaption{\rm Sources not detected in VCS6 VLBA observations\label{t:nondetected}}
\tabletypesize{\scriptsize}
\tablehead{
\vspace{2.0ex} \\
   \multicolumn{2}{c}{Source name}          &
   \multicolumn{2}{c}{J2000.0 Coordinates}  &
   \vspace{0.5ex} \\
   \colhead{IVS}      &
   \colhead{IAU}      &
   \colhead{Right ascension} &
   \colhead{Declination} &
   \colhead{List} 
   \vspace{0.5ex} \\
   \colhead{(1)}    &
   \colhead{(2)}    &
   \colhead{(3)}    &
   \colhead{(4)}    &
   \colhead{(5)}     
   }
\startdata
%
\objectname{0123{\tt-}015} & \objectname{J0126{\tt-}0118}     & 01 26 04.77 & {\tt-}01 18 16.2 &  v \vspace{0.5ex} \\
\objectname{0428{\tt-}252} & \objectname{J0430{\tt-}2507}     & 04 30 16.05 & {\tt-}25 07 38.6 &  v \vspace{0.5ex} \\
\objectname{0741{\tt-}182} & \objectname{J0743{\tt-}1825}     & 07 43 34.76 & {\tt-}18 25 03.1 &  v \vspace{0.5ex} \\
\objectname{0829{\tt+}140} & \objectname{J0831{\tt+}1353}     & 08 31 59.12 & {\tt+}13 53 15.4 &  v \vspace{0.5ex} \\
\objectname{0916{\tt-}219} & \objectname{J0918{\tt-}2206}     & 09 18 30.27 & {\tt-}22 06 55.3 &  v \vspace{0.5ex}
\enddata
  \tablecomments{ Table~\ref{t:nondetected} is presented in its entirety 
                  in the electronic edition of the Astronomical Journal. 
                  A portion is shown here for guidance regarding its form and 
                  contents. Units of right ascension are hours, minutes and 
                  seconds; units of declination are degrees, minutes and 
                  seconds. The J2000 source positions are taken from the NVSS
                  survey \citep{nvss}, they were used for VCS6 VLBA observing 
                  and correlation.
                }
\end{deluxetable}

  Including VCS6 observations raised the total number of calibrators with
declination $\delta>-40\degr$ to 2853. The sky calibrator density for 
different radii of a search circle for declination $\delta>-40\degr$ 
is presented in Figure~\ref{f:prob}.

  In addition to these tables, the html version of this catalogue is posted
on the Web\footnote{\url{http://vlbi.gsfc.nasa.gov/vcs6}}. For each source
it has 8 links: to a pair of postscript maps of the source at X and S-band;
to a pair of plots of correlated flux density as a function of the length
of the baseline vector projection to the $uv$ plane; to a pair of fits 
files with CLEAN components of naturally weighted source images; and to 
a pair of fits files with calibrated $uv$ data. The coordinates and the plots 
are also accessible from the NRAO VLBA Calibration Search 
web-page\footnote{\url{http://www.vlba.nrao.edu/astro/calib}}. 
Figure~\ref{f:images} presents examples of naturally weighted contour CLEAN 
images as well as correlated flux density versus projected spacing dependence 
for three sources from the lists `r', `i', and `v', respectively: 
\objectname{J0612+6225}, \objectname{J2237+4216}, and \objectname{J1233+8054}.

\section{Summary}\label{s:summary}

   The VCS6 Survey has added 169 new compact radio sources not previously 
observed with VLBI and significantly improved coordinates of 95 other objects.
Among the new sources, 103 objects turned out to be suitable as phase
referencing calibrators and as target sources for geodetic applications. 
Their coordinates have position accuracy better than 5~mas.

\acknowledgments

\ifpre 
  {The National Radio Astronomy Observatory is a
   facility of the National Science Foundation operated under cooperative
   agreement by Associated Universities, Inc.} 
\else
  \facility[NRAO(VLBA)]{The National Radio Astronomy Observatory is a
   facility of the National Science Foundation operated under cooperative
   agreement by Associated Universities, Inc.} 
\fi
We thank the staff of the
VLBA for carrying these observations in their usual efficient manner.
This work was done while L.~Petrov and D.~Gordon worked for NVI, Inc. 
under NASA contract NAS5--01127. Y.~Y.~Kovalev is a Research Fellow of the 
Alexander von Humboldt Foundation. \mbox{RATAN--600} observations were partly 
supported by the Russian Foundation for Basic Research (projects 01--02--16812
and 08--02--00545). The authors made use of the database CATS \citep{cats} 
of the Special Astrophysical Observatory. This research has made use 
of the NASA/IPAC Extragalactic Database (NED) which is operated by the 
Jet Propulsion Laboratory, California Institute of Technology, under 
contract with the National Aeronautics and Space Administration.

\end{document}